\documentstyle[12pt]{article}
\newcommand{\NP}[1]{Nucl. \ Phys.}
\newcommand{\PL}[1]{Phys. \ Lett.}
\newcommand{\PRL}[1]{Phys.\ Rev.\ Lett. }
\newcommand{\AP}[1]{Ann.\ Phys. }
\newcommand{\IM}[1]{Inv.\ Math. }
\newcommand{\JMP}[1]{ J.\ Math.\ Phys. }
\newcommand{\RMP}[1]{Rev.\ Mod.\ Phys.}
\newcommand{\JDG}[1]{ J.\ Diff.\ Geom. }
\newcommand{\PTP}[1]{ Prog.\ Theor.\ Phys. }
\newcommand{\SPTP}[1]{Suppl.\ Prog.\ Theor.\ Phys. }
\newcommand{\PR}[1]{Phys.\ Rev. }
\newcommand{\PREP }[1]{Phys.\ Reports }
\newcommand{\NC}[1]{Nuovo \ Cim. }
\newcommand{\NCL }[1]{Nuovo\ Cim.\ Lett. }
\newcommand{\CMP}[1]{Commun.\ Math.\ Phys. }
\newcommand{\TMF}[1]{Theor.\ Math.\ Phys. }
\newcommand{\CQG}[1]{Class.\ Quant.\ Grav. }
\newcommand{\FAA}[1]{Funct. Analys. Appl. }
\newcommand{\JP}[1]{J.\ Phys. }
\newcommand{\JA }[1]{ J. Algebra }
\newcommand{\JFA}[1] {J. Funct. Anal. }
\newcommand{\MPL}[1] { Mod. Phys. Lett. }
\newcommand{\IJMP}[1] { Int. J. Mod. Phys. }
\newcommand{\RMAP}[1] { Rep.\ Math.\ Phys. }

\newcommand{\ba}{\begin{array}}
\newcommand{\ea}{\end{array}}
\newcommand{\eps}{\epsilon}
\newcommand{\veps}{\varepsilon}
\newcommand{\ra}{\rightarrow}

\newcommand{\be}{\begin{equation}}
\newcommand{\ee}{\end{equation}}
\newcommand{\bea}{\begin{eqnarray}}
\newcommand{\eea}{\end{eqnarray}}

\begin{document}
\centerline{\Large \bf Vanishing of Cosmological Constant}
 
 \bigskip
\centerline{\Large \bf in Dual Supergravity
\footnote{Research supported by the
 RFFI Grant 95-02-05822a} }

\vspace*{ 1cm}

\centerline{\bf
M.V.Terentiev,  K.N.Zyablyuk }

\centerline{$  Institute \ \ of \ \ Theoretical \ \
 and \ \  Experimental \ \ Physics $ }

\bigskip
\begin{abstract}

The vacuun configuration of dual supergravity in ten dimensions with one-loop
fivebrane corrections is analyzed.  It is shown that
the compactification of this theory with rather general conditions
  to  six dimensional space  leads to  zero value of
cosmological constant.

\end{abstract}

\section{Introduction}
The cosmological constant $C$ can be  defined as a vacuum expectation value
(VEV) of an effective lagrangian:
\be
C= <{\cal L}_{eff}>
\ee
It  vanishes $(C=0)$ in any theory with global
supersymmetry, - that is a direct consequence of a supersymmetry algebra.
The situation is not so unambiguous in a supergravity, where in general
$ C \ne 0 $. But $C=0$ for the heterotic superstring in ten dimensions
(D=10).
So, one may expect that the Type I (N=1, D=10) supergravity considered as
a field-theory limit of a heterotic superstring also leads to $C=0$.
The same must be true for the dual N=1, D=10 supergravity considered as a
field-theory limit of a fivebrane \cite{S1}, \cite{D}, \cite{DL}
 because this theory
can be obtained from a Type I supergravity by dual transformation of
the axionic field.

It is not evident, that the condition $C=0$ will persist in a
 compactification for lower dimensions because the part of the
supersymmetry can be lost and vacuum properties are rather specific
in the process.

We demonstrate in the present paper, that rather general and realistic
compactification procedure from the space $M_{10}$ to the $M_6$ leads to
vanishing of the cosmological constant.
 (Here $M_D $ is a $D$-dimensional
space-time with Minkowsky signature). Really the case is considered:
\be
                                                                 \label{2}
M_{10} \ra M_{6} \otimes E_4
\ee
where the compact Eucledian D=4 space $E_4$. We require that vacuum
configuration leaves only halph of the D=10 supersymmetry unbroken and
leads to chiral theory in D=6. This assumption greatly 
simplifies the equations for vacuum configuration and uniquely 
fixes the topological structure of $E_4$ manifold. It follows, that
$E_4$ is  related to the  $K_3$-space 
(see \cite{P} for referencies on $K_3$-space).

We concentrate on the dual supergravity because the supersymmetric
lagrangian for this theory is constructed including terms of the next
order in the string-tension parameter $\alpha'$
 (see \cite{STZ1}, \cite{STZ2}).

\section{Lagrangian}

The lagrangian of dual supergravity in $M_{10}$ is equal to:
\be
                                                                    \label{3}
{\cal L} =
{\cal L}^{(gauge)}+ {\cal L}^{(grav)}
\ee
where ${\cal L}^{(gauge)}$ and ${\cal L}^{(grav)} $ are lagrangians for
gauge-matter and for supergravity multiplet. The term
${\cal L}^{(gauge)}$ takes the form  \cite{CH}:
\be
                                                                 \label{4}
E^{-1}\,{\cal L}^{(gauge)}={1\over g^2}\, tr\, \left[
{1\over4}\,{\cal F}_{AB}\,{\cal F}^{AB}-
{1\over{8 \cdot 6!}}\,{\varepsilon}^{A_1 \ldots A_{10}}\,C_{A_1 \ldots A_6}
\,{\cal F}_{A_7A_8}\,
{\cal F}_{A_9A_{10}} \right]
\ee
Here  $C_{A_1 \ldots A_6}$ is an axionic potential, ${\cal F}_{AB} $ is a
 gauge field which is in the algebra of internal symmetry group $G$,
 ${\cal F}_{AB} = E_A^ME_B^N \, F_{MN}, $ where $E_M^A$ is the veilbein in
$M_{10}$.

 One must consider $G=SO(32)$ or $G=E_8\times E_8$ and $ 4\,g^2 =-1/\alpha'$
 as it follows from
the superstring consideration \cite{GS}. The symbol $tr$ in (\ref{4})
means the trace in a vectorial representation of $SO(32)$. It can be
changed to
$ (1/30)Tr $, where $Tr$ means the trace in the adjoint representation of
$E_8\times E_8$ or $SO(32)$.

We consider   only  bosonic terms in  the lagrangian
 (\ref{3}) because we are interested in  the
 vacuum configuration. Notations correspond in general
 to \cite{STZ1} with some differencies which are or
self-evident or explained in
the text.   In particular, the following index notations are used:
 $A,B,C,\ldots$,
 $  a,b,c,\ldots $ and   $\alpha, \beta, \gamma, \ldots $  are flat
 indices  respectively in $M_{10}$, $M_6$ and $E_4$;
 $  M,N,P,\ldots $,
$ m,n,p,\ldots$ and $ \mu, \nu, \lambda, \ldots $ are corresponding world
indices; $Z^M =(x^\mu, y^m)$ is the coordinate in $M_{10}$, $x,\,y$ are
coordinates in $M_6$ and $E_4$ respectively.

We present the gravity part of the lagrangian as an expansion in $\alpha'$:
\be
                                                                   \label{5}
{\cal L}^{(grav)}= {\cal L}^{(grav)}_0  +\alpha' {\cal L}^{(grav)}_1
\ee
where $ {\cal L}^{(grav)}_0$ is equal to \cite{CH}
(see \cite{STZ2} for further references on the subject):
\be
                                                                   \label{6}
 E^{-1}\,{\cal L}^{(grav)}_0 = \phi\,\left({\cal R}-
{1\over 12}\,{\tilde M}_{ABC}^2 \right)
\ee
Here ${\cal R} $ is the curvature scalar, $\phi$ is the dilatonic
field, ${\tilde M}_{ABC} $ is defined by:
\be
                                                                    \label{7}
{\tilde M}_{ABC} = {1 \over 7!}{\veps_{ABC}}^{A_1\ldots A_7}
M_{A_1\ldots A_7}
\ee
where $M_{N_1\ldots N_7} = 7\,\partial_{[N_1}C_{N_2 \ldots N_7]}$
is the axionic field-strength.

The result for  ${\cal L}^{(grav)}_1 $ was obtained in \cite{STZ1},
\cite{STZ2}  in the form:
$$ {\cal L}^{(grav)}_1 =  2\,{\cal R}^2_{AB} -{\cal R}_{ABCD}^2 +
{1\over 2\cdot 6!}
\varepsilon^{ABCDF_1\ldots F_6}\,{{\cal R}_{AB}}^{IJ}
 {\cal R}_{CDIJ}C_{F_1\ldots F_6} - $$
$$ -{1\over 2}\,{\cal R}^{AB}({\tilde M}^2)_{AB}
 -{1\over 6}\, {\tilde M}^{ABC}D_F^2{\tilde M}_{ABC} + $$
    \be
                                                               \label{8}
+ {1\over 2}\, {\tilde M}^{ABC;D}({\tilde M}^2)_{ABCD}
  -{1\over 24}\, ({\tilde M}^2)_{ABCD}({\tilde M}^2)_{ACBD}
\ee
Here ${\cal R}_{ABCD} $ is the curvature tensor, ${\cal R}_{AB}$ is the
Ricci tensor, $;B$ means the covariant derivative $D_B $.
 The following notations are introduced here and below:

$$ {\tilde M}^2 =  ({\tilde M}_{ABC})^2, \ \
 ({\tilde M}^2)_{AB} = {{\tilde M}_A}^{CD} {\tilde M}_{BCD} $$
$$ ({\tilde M}^2)_{ABCD}=
 {{\tilde M}_{AB}}{}^F {\tilde M}_{CDF}, \ \
({\tilde M}^3)_{ABC} = {{\tilde M}_A}{}^{IJ} {{\tilde M}_{BJ}}{}^K
{\tilde M}_{CKI} $$
The ${\tilde M}_{ABC} $ field is connected by the dual transformation
with the 3-form axionic field of standard Type I supergravity (see below).

\section{Vacuum Configuration}

The most general anzatz for VEV of the veilbein in $M_{10} $, which
corresponds to the compactification according to eq.(\ref{2}),
 takes the form:
\be
                                                                  \label{9}
{E_M}^A = \left(
\ba{cc}
e^{2\,\xi(y)}\,\delta_\mu^\alpha  &   0 \\
0         & e^{-2\,\xi(y)}\, {e_m}^a(y)
\ea
\right)
\ee
where $ {{\tilde e}_m}^a \equiv \exp(-2\,\xi)\, {e_m}^a$
 is the veilbein in $E_4\ $  (the factor $ \exp(-2\,\xi) $ is extracted for
future convenience),  $\xi $
is an arbitrary function. The analogous ansatz was considered
 \cite{S2} in the study of compactification
 scheme $ M_{10}\ra M_4 \otimes E_6 $ with only partial account of $\alpha'$
corrections.

Only the gauge-field component ${\cal F}_{ab} =\exp(2\xi) \, F_{ab} $
 survive
 in the vacuum configuration ( $F_{ab} $ $ = e_a{}^m\,e_b{}^n\,F_{mn}$ ).

The folowing VEV's  of curvature tensor componets  survive
(${\cal R} =d\omega + \omega \wedge \omega $
 where $\omega $ is the connection related with $E_M{}^A$):

$${{\cal R}_{\alpha\beta}}^{\gamma \delta}=
2\,e^{4\xi}\,\delta^{\,\gamma}_{[\alpha}
\delta^{\,\delta}_{\beta]}\,\xi^f\xi_f $$
$$ {{\cal R}_{\alpha b}}^{\gamma c}=e^{4\xi} \delta_\alpha^\gamma
\,({\xi_b}^c + 5\,\xi_b\xi^c -2\, \delta_b^c\, \xi^f\xi_f) $$
\be
                                                                 \label{10}
{{\cal R}_{ab}}^{cd}=e^{4\xi}\,(R_{ab}{}^{cd} -8\,
\delta_{[a}^{}{[c} \,\xi_{b]}{}^{d]} -
16\,\delta_{[a}{}^{[c} \,\xi_{b]}\xi^{d]} +8\,\delta_{[a}^{\,c}\delta_{b]}^d\,
\xi^f\xi_f)
\ee
where $R_{mnab} $ is the curvature defined in terms of ${e_m}^a$,
$\xi_b =\nabla_b\,\xi= e_b{}^m \,\partial_m \,\xi,$ etc.

Let us start now to study equations  defining vacuum confi\-guration:
 $ <\delta_Q \, \Phi >= 0, $
 where $\Phi $ is some field but $\delta_Q$
 is a super\-sym\-metry transfor\-mation.
 When $\Phi$ is a boson, such an
equation satisfied identically. It has nontrivial content for
 $\Phi =\psi_A,$ $\chi,$  $\lambda $, i.e. for
gravitino, dilatino and gaugino fields respectively.

 We get \cite{STZ3} (see also \cite{BBLPT}, \cite{AFRR} where another
parametrization is used):

\be
                                                                \label{11}
<\delta_Q\psi_A> = \eps_{;A} +{1\over 144}
 \left( 3\,{\tilde M}_{BCD}\Gamma^{BCD}\Gamma_A+
\Gamma_A{\tilde M}_{BCD}\Gamma^{BCD}\right)\eps =0
\ee
\be
                                                                   \label{12}
<\delta_Q \chi> = {1\over 2}\partial_A \phi \, \Gamma^A \eps -
\left( {\phi\over 36}\,{\tilde M}_{ABC}\Gamma^{ABC} -\alpha'\,
A_{ABC}\Gamma^{ABC}\right)\eps =0
\ee
\be
                                                                  \label{13}
<\delta_Q \lambda> = {1\over 4}{\cal F}_{AB}\Gamma^{AB} \eps =0
\ee
Here $\eps $ is a 32-component Dirac spinor, - the parameter of
supersymmetry transformation. It is subjected to the Majorana-Weyl
condition: $ \eps_c = {\bar\eps}\  $, $\eps = \Gamma \eps $, where
$ \Gamma $ is the chirality matrix (see below), $\eps_c$ is
the charge-congugated spinor.
 We suppose that $\eps $ is independent on
the coordinates in $M_6$: $\eps = \eps(y) $, that corresponds to the
symmetry of vacuum configuration. All the fields in the r.h.s. of
 (\ref{11})-(\ref{13}) depend only on $y$.

The 3-form field $A_{ABC} $ in eq.(\ref{12}) is equal to \cite{STZ1},
\cite{STZ2}:

$$ A_{ABC} = -{1\over 18} \Box {\tilde M}_{ABC} +
{7\over 36}({\tilde M}^2_{D[ABC]}){}^{;D} +{1\over 36} {\tilde M}_{DE[A;B}
{{\tilde M}_{C]}}{}^{DE}- $$
$$- {5\over 8\cdot 243}{\tilde M}^2{\tilde M}_{ABC}+
{5\over 8\cdot 27} {\tilde M}^2_{D[A}{{\tilde M}_{BC]}}{}^D
 -{5\over 4\cdot 27}{\tilde M}^3_{ABC}-$$
\be
                                                                   \label{14}
-{1\over 4\cdot 972}{\veps_{ABC}}^{DEFGHIJ}{\tilde M}_{DEF}
({\tilde M}_{HIJ;G} +{\tilde M}^2_{GHIJ})
\ee

The following representation of $\Gamma $-matrices $M_{10}$ is convenient:
$$ \Gamma^\alpha = \gamma^\alpha \otimes {\cal T}, \ \ \
\alpha = 0,1,\ldots ,5 $$
\be
                                                                 \label{15}
\Gamma^{a+5} = I \otimes {\cal T}^a, \ \ \ a=1,2,3,4
\ee
where $\gamma^\alpha $ are $8\times 8 $ Dirac matrices in $M_6$, but
${\cal T}^a $ are $4\times 4$ Dirac matrices in $E_4$. The chirality
matrices are:
\be
                                                                 \label{16}
\gamma = \gamma^0\gamma^1\ldots\gamma^5, \ \ \
{\cal T} = {\cal T}^1{\cal T}^2{\cal T}^3{\cal T}^4
\ee
Then
\be
\label{17}
\Gamma =\Gamma^0\Gamma^1 \ldots \Gamma^9 = \gamma \otimes {\cal T}, \ \ \
(\Gamma)^2 =(\gamma)^2 =({\cal T})^2 =1
\ee
We use also the standard notation: $ \Gamma_{A_1\ldots A_k}
= \Gamma_{[A_1}\Gamma_{A_2} \ldots \Gamma_{A_k]}. $

We begin our study from  eq.(\ref{11}). For $A=\alpha $  it
reduces to:
\be
\label{18}
\gamma_\alpha \gamma \otimes \left( e^{2\xi}\,\xi_a\,{\cal T}^a +
{1\over 36} {\tilde M}_{abc}\,{\cal T}^{abc} \right)\eps=0
\ee
Let us suppose, that $\eps$ has definite chirality in $E_4$:
\be
                                                               \label{19}
{\cal T}\eps = \nu \eps, \ \ \mbox{where} \ \nu=1 \ \mbox{or} \  -1
\ee
It means that we keep only "one-halph" of the supersymmetry in $M_{10} $.
Using the relation:
$$ {\cal T}^{abc} = - \veps^{abcd}\,{\cal T}_d \,{\cal T}    $$
where $\veps^{abcd} $ is a completely antisymmetric tensor
($\veps^{1234} =1$), we get the solution of (\ref{18}) in the form:
\be
                                                                  \label{20}
{\tilde M}_{abc} = 6\,\nu \,e^{2\,\xi}\,\veps_{abcd}\, \xi^d
\ee
It means that the only nonzero component of the axion potential with all
flat $M_6$ indices appears to be a constant:
\be
                                                                 \label{21}
C_{\beta_1\ldots \beta_6} = -\nu\, \veps_{\beta_1\cdots \beta_6}
\ee
where $\veps^{\beta_1 \ldots \beta_6} $ is a
 completely antisymmetric tensor ($\veps^{01\ldots 6} =1$).

Having all this, one can reduce eq.(\ref{11}) with $A=a $ to the
following:
\be
                                                                    \label{22}
\nabla_a \eps' = 0, \ \ \mbox{where} \ \ \eps'= e^{-{\xi/2}}\,\eps
\ee
Equation $[\nabla_a, \nabla_b]\eps'= 0 $ leads to:
\be
                                                                \label{23}
R_{abcd}\,{\cal T}^{cd}\, \eps=0
\ee
Eq. (\ref{23}) is fulfilled for arbitrary chiral $\eps$ if the
 curvature defined by the veilbein ${e_m}^a$ is (anti)selfdual for the
second pair of indices:
\be
                                                                  \label{24}
R_{abcd} = {\nu \over 2}\,{\veps_{cd}}^{ef}\,R_{abef}
\ee
This equation immediately follows from (\ref{23}) because ${\cal T}^{ab}=
-(1/2)\veps^{abcd}\,{\cal T}_{cd}\,{\cal T} $.
  It follows from (\ref{24}) with the help of  simmetry properties of
the curvature tensor  that curvature is
 (anti)selfdual for the first pair of indices too. The compact 4-dimensional
space with the (anti) selfdual curvature is the $K_3$ space. It means that the
space ${\tilde E}_4$ which is defined by the veilbein ${e_m}^a$ is the
$K_3 $-space. This result follows from the ansatz (\ref{9}) with the constraint
(\ref{10}).

Now we turn to the eq.(\ref{13}).
 It is analogous to eq.(\ref{23}), so it
leads to the selfduality of the gauge-field ${\cal F}_{ab}$:
 \be
                                                               \label{25}
{\cal F}_{ab} ={\nu \over 2}\, {\eps_{ab}}^{cd}\,{\cal F}_{cd}
\ee

We are left with eq.(\ref{12}) which is the most complicated one.
Taking into account eqs. (\ref{10}), (\ref{20}) one is able
 to present the $A_{abc} $-tensor  from eq.(\ref{14}) in
a rather simple form:
\be
                                                                  \label{26}
A_{abc} = -{\nu\over3}\, \veps_{abcd}
\left[e^{6\xi}\,({\xi^f}_f +
\xi^f\xi_f)\right]^{;d}.
\ee
where $;d$ means the  covariant derivative $\nabla_d$ defined by the
veilbein $e_m{}^a$.

Expresion (\ref{26}) allows us to drop one derivative
 in eq.(\ref{12}) reducing the
order of equation from third to  second. Then it follows immediately
from eq.(\ref{13}):
\be
                                                                  \label{27}
e^{2\xi}\, \phi +4\,\alpha'\, e^{6\xi}\,({\xi^f}_f +\xi^f\xi_f) =C_0
\ee
where $C_0$ is an arbitrary constant. This equation connects the dilaton VEV
with  function
$\xi(y)$ introduced by the anzatz (\ref{9}).

\section{Equations of Motion}

We examine here whether equations of motion impose some additional
constraints on the parameters of vacuum configuration. Equations
for $M_{10}$ supergravity obtained in the lowest and next order in
$\alpha' $  are used (see \cite{STZ1}, \cite{STZ2}), see also  \cite{P}
where another parametrization was considered):

One can easily see that gauge-field equations of motions are fulfilled
because of selfduality condition and Bianchi Identity $\nabla_{[a}F_{bc]}=0$.

The dilaton equation of motion can be presented in the form:

$$\Box \phi +{1\over12}\phi {\tilde M}^2_{ABC}+
{1\over 12\,g^2}tr\,({\cal F}_{AB})^2
-\alpha'\left[ -{2\over 3}({\cal R}_{AB})^2
 +{1\over3}({\cal R}_{ABCD})^2 -\right.$$
$$-{1\over 6}{\cal R}^{AB}{\tilde M}^2_{AB}
  -{1\over18}{\tilde M}^{ABC}\Box {\tilde M}_{ABC}+
{1\over3}{\tilde M}^{ABC;D}{\tilde M}^2_{ABCD}-$$
\be
                                                                 \label{28}
\left. -{1\over24} {\tilde M}^2_{ABCD} {\tilde M}^{2 \ ACBD}
 -{1\over12} \Box {\tilde M}^2 +
{1\over6} ({\tilde M}^2_{AB})^{;AB} \right] =0
\ee
where $\Box = D_B^2$.

Calculating all the terms in eq.(\ref{28}) with the help of relations
obtained before and using eq.(\ref{27}), one obtains the following
result:
\be
                                                                  \label{29}
{\left( e^{-6\xi}C_0 -12\,\alpha'\,{\xi_a}^a \right)^{;b}}_{;b}+
{1\over 4g^2}\, tr\,(F_{ab})^2 -\alpha' (R_{abcd})^2 =0
\ee
where $F_{ab} = e_a{}^m e_b{}^n F_{mn}\ $, $R_{abcd}=e_a{}^m e_b{}^n
R_{mncd} $ .

One can expect (cf. \cite{CHSW})
that VEV's  of supersymmetry transformations
 (\ref{11})-(\ref{13})
provide the complete information on the vacuum configuration, i.e.
they are equivalent to equations of motion. In such a case,
one must be able to derive eq.(\ref{29}) starting immediately from
(\ref{11})-(\ref{13}). But up to now  we were not able to do this.

One gets immediately from (\ref{29}) the
topological constraint \cite{W} (in form notations):
\be
                                                                    \label{30}
\int_{K_3} (tr\,F\wedge F -tr\,R\wedge R) =0
\ee
Here $tr$ is calculated over indices of vectorial representation
of corrsponding group, i.e $tr\,R\wedge R)= R_{ab}\wedge R^{ba} $.
 The integral is fulfilled over the compact space with the
veilbein $e_m^a $ (i.e the $K_3$-space).

Now we turn to the equation of motion for the $C_{A_1\ldots A_6}$-field.
One can write it in the form \cite{STZ1}:

\be
                                                                   \label{31}
H_{[ABC;D]}+ 3\,\alpha' \left( tr\, {\cal F}_{[AB}{\cal F}_{CD]}-
{\cal R}_{[AB}{}^{EF}{\cal R}_{CD]FE} \right) =0
\ee
where

$$ H_{ABC} =\phi \, {\tilde M}_{ABC} -2\,\alpha'\Bigl(-\Box {\tilde M}_{ABC}
 +3\,({\tilde M}^2_{D[ABC]})^{;D} + $$
\be
                                                                  \label{32}
+{3\over 2}{\tilde M}_{DF[A;B}{\tilde M}_{C]}{}^{DF}
 -3\,{\cal R}_{D[A} {\tilde M}_{BC]}{}^D -
{1\over 2}{\tilde M}^3_{[ABC]}\Bigr)
\ee
Only the $H_{abc}$-component survive in the vacuum configuration.
The calculation of this component is similar to that performed
for $A_{abc}$.  The result is:
\be
                                                                  \label{33}
H_{abc} = 6\,\nu\,\eps_{abcd}\,(C_0 \,\xi^d +2\,\alpha'\, e^{6\xi}\,
{\xi^{df}}_f)
\ee
Then, one can check with the help of selfduality conditions
 that eq.(\ref{31}) is equivalent to eq.(\ref{29}), i.e.
no new constrains are produced.

So we obtain, that eq.(\ref{29}) is
the VEV of usual $M_{10}$ supergravity Bianchi Identity:
\be
                                                                   \label{34}
dH' = 2\alpha'(-tr \, F\wedge  F +tr \,  R\wedge  R)
\ee
where $H'_{abc}$ is interpreted as a VEV of  axionic  field-strength of
 usual Type I supergravity. It takes the form
\be
                                                                  \label{35}
H'_{abc} =\nu\,\eps_{abcd}\,H^{;d}
\ee
where $H(y)$ is a scalar field. It follows from  (\ref{29}):
\be
                                                                 \label{36}
{(H-e^{-6\xi}\,C_0 +12\, \alpha'\, {\xi^a}_a)^{;b}}_{;b}=0
\ee
One can easily find the relation between $H'_{abc}$ and $H_{abc} $-field in
(\ref{33}).

A long study of a rather complicated graviton equation of
motion ($\phi {\cal R}_{AB} + \ldots =0 $) does not produce  additional
constraints for vacuum configuration.

\section{Zero Cosmological Constant}

In this section we demonstrate that the action VEV  vanishes.
The internal space assumed to be closed, so surface integrals
do not give any contribution.

Using eqs.(\ref{10}) and (\ref{20}) one gets immediately:
\be
                                                                    \label{37}
<\,{\cal L}_0^{(grav)}\,> =0
\ee
where ${\cal L}_0^{(grav)}$ is defined in eq.(\ref{6}).

The gauge-matter part of the lagrangian (\ref{3}) can be transformed to
the form:
\be
                                                                  \label{38}
<\,E^{-1}{\cal L}^{(gauge)}\,> ={1\over 4g^2}\,tr\,({\cal F}_{ab}^2-
{\nu\over2}\,\eps^{abcd}{\cal F}_{ab}{\cal F}_{cd})
\ee
where eq.(\ref{21}) was used. Then, the selfduality condition (\ref{25})
leads immediately to:
\be
                                                                  \label{39}
<\,{\cal L}^{(gauge)}\,>=0
\ee
The most difficult term in (\ref{3}) is the ${\cal L}_1^{(grav)}$.
Using (\ref{8}) and  relations obtained before we are able
 to transform  this term to the form of complete derivative:
\be
                                                                \label{40}
<\,E^{-1}{\cal L}_1^{(grav)}\,> =
\nabla_a \,\left[ e^{6\xi}\,(12\,\xi^{ab}\xi_b-3\,\xi^a{\xi_b}^b -
12\,\xi^a\xi^b\xi_b)\right]
\ee
 So, one can put:
\be
                                                                \label{41}
<\,\int \,d^{10}Z\,{\cal L}_1^{(grav)}\,>=0
\ee
Then
\be
                                                                \label{42}
C = <\,\int \,d^{10}Z\,  ({\cal L}^{(gauge)}+{\cal L}_0^{(grav)}+
\alpha'\, {\cal L}_1^{(grav)})\,> =0.
\ee
We conclude, that no cosmological constant is generated.


\begin{thebibliography}{99}
\bibitem{S1}\ A.Strominger, ~~Nucl. Phys. {\bf B 343} (1990) 167.
\bibitem{D}\ M.J.Duff, \ \ Class. Quant. Grav. {\bf 5} (1988) 189.
\bibitem{DL}\ M.J.Duff, J.X.Lu  \ \ Phys. Rev. Lett.   {\bf 66} (1988) 1402, \\
Class. Quant. Grav/ {\bf 9} (1991) 1.
\bibitem{PAGE} \ D.Page, \ \ Phys. Lett. {\bf 80B} (1978) 55.
\bibitem{STZ1} \ N.A.Saulina,  M.V.Terentiev, K.N.Zyablyuk, \\
~~\PL~~{\bf B 366} (1966) 134
\bibitem{STZ2} \ N.A.Saulina,  M.V.Terentiev, K.N.Zyablyuk, \\
 Nucl. Phys. (submitted to publication),  hep-th XXXXXXXXX
\bibitem{CH} \ A.Chamseddine, ~~Phys. Rev. {D 24} (1981) 3065.
 \bibitem{GS}\ M.Green, J.Schwarz,~~\PL~~{\bf B 149} (1984) 117.
\bibitem{S2} \ A.Strominger, ~~Nucl. Phys. {\bf B 274} (1986) 253.
\bibitem{STZ3} \ N.A.Saulina,  M.V.Terentiev, K.N.Zyablyuk, \\
~~Yad. Fiz.~~{\bf 59} (1966) 163.
\bibitem{BRW} \ E.Bergshoeff, M.De Roo, B. De Witt,
~~\NP~~ {\bf B 217} (1983) 489.
\bibitem{BBLPT} \  L.Bonora, M.Bregola, K.Lechner, P.Pasti, M.Tonin,  \\
 ~~\NP ~~{\bf B 296} (1988) 877.
 \bibitem{AFRR}\ R.D'Auria, P.Fre,
M.Raciti, F.Riva, ~~Intern. J. Mod. Phys.  {\bf A 3}\ (1988)\ 953.
\bibitem{P} \ I.Pesando, ~~Phys. Lett. {\bf B 272} (1991) 45, \\
Class. Quantum Grav. {\bf 9} (1992) 823.
\bibitem{CHSW} \ P.Candelas, G.Horowitz, A.Strominger, E.Witten, \\
Nucl. Phys. {\bf B 258} (1985) 46.
\bibitem{W} \ E.Witten, ~~Phys. Lett. {\bf B 149} (1984) 351.

\end{thebibliography}
\end{document}